\documentstyle[emulateapj,apjfonts,psfig,natbib,amsmath]{article}

\begin{document}
\def\cfa{1}
\def\uva{2}

\title{Radio Observations Reveal Unusual Circumstellar Environments
  for Some Type Ibc Supernova Progenitors}

\author{Sarah Wellons\altaffilmark{\cfa}, Alicia M. Soderberg\altaffilmark{\cfa}, and Roger A. Chevalier\altaffilmark{\uva}}
\altaffiltext{\cfa}{Harvard-Smithsonian Center for Astrophysics, 60 Garden St., Cambridge, MA 02138, USA}
\altaffiltext{\uva}{University of Virginia, Astronomy Department, Charlottesville, VA 22904, USA}

\begin{abstract}
We present extensive radio observations of the nearby Type Ibc
supernovae (SNe Ibc) 2004cc, 2004dk, and 2004gq spanning $\Delta
t\approx 8-1900$ days after explosion.  Using a dynamical model
developed for synchrotron emission from a slightly decelerated
shockwave, we estimate the velocity and energy of the fastest ejecta
and the density profile of the circumstellar medium.  The shockwaves
of all three supernovae are characterized by non-relativistic
velocities of $\overline{v}\approx (0.1-25)c$ and associated energies
of $E\approx (2-10)\times 10^{47}$ erg, in line with the expectations
for a typical homologous explosion.  Smooth circumstellar density
profiles are indicated by the early radio data and we estimate the
progenitor mass loss rates to be $\dot{M}\approx (0.6-13)\times
10^{-5}~\rm M_{\odot}~yr^{-1}$ (wind velocity, $v_w=10^3~\rm
km~s^{-1}$).  These estimates approach the saturation limit
($\dot{M}\approx 10^{-4}~\rm M_{\odot}~yr^{-1}$) for line-driven winds
from Wolf-Rayet stars, the favored progenitors of SNe Ibc
including those associated with long-duration gamma-ray bursts.
Intriguingly,  at later epochs all three supernovae show evidence
for abrupt radio variability that we attribute to large density
modulations (factor of $\sim 3-6$) at circumstellar radii of $r\approx
(1-50)\times 10^{16}$ cm.  If due to variable mass loss, these
modulations are associated with progenitor activity on a
timescale of $\sim 10-100$ years before explosion.  We consider these results in the context of
variable mass loss mechanisms including wind clumping,
metallicity-independent continuum-driven ejections, and binary-induced
modulations.  It may also
be possible that the SN shockwaves are dynamically interacting with 
wind termination shocks, however, this requires the environment to 
be highly pressurized and/or the progenitor to be rapidly rotating 
prior to explosion.  The proximity of the density modulations to the
explosion sites may suggest a synchronization between unusual progenitor mass loss
and the SN explosion, reminiscent of Type IIn
supernovae.  This study underscores the utility of radio observations for tracing the final
evolutionary stage(s) of SN progenitor systems.
\end{abstract}
\keywords{supernovae}

\section{Introduction}
\label{sec:intro}

Type Ibc supernovae (SNe Ibc) are commonly understood to mark the
death of massive stars, $M\gtrsim 10~M_{\odot}$. A relatively rare
sub-group, SNe Ibc represent just 10-20\% of all local SN discoveries
\citep{cet99,sma09,llc+10}. They are distinguished from other
core-collapse SNe by the lack of hydrogen features in their optical
spectra (see \citealt{Filippenko1997} for a review) indicating that
their massive hydrogen envelopes are removed prior to explosion
\citep{elias85,alex85,Wheeler&Levreault1985,uomoto85,clo86,whw02}.
The lack of helium absorption features may further divide the sample
into SNe Ic (helium-poor) and SNe Ib (He-rich; \citealt{mfl+01} but
see \citealt{hmp+02}).

The physical mechanism and timescale on which SNe Ibc progenitors lose
their envelopes, however, remains an open question.  In a favored
progenitor model, the metallicity-dependent (and often clumpy)
line-driven wind of an isolated Wolf-Rayet (WR) star expels its own
hydrogen envelope \citep{Begelman&Sarazin1986,
Woosley_et_al1995}. However, as argued by
\citet{Smith&Owocki2006}, metallicity-independent continuum-driven winds 
and/or hydrodynamic eruptions may play a key role in the formation of
Wolf-Rayet stars.  A binary mechanism such as Roche lobe overflow,
accretion or a common envelope phase could also remove the envelope
of a massive progenitor \citep{Wheeler&Levreault1985,
Podsi_et_al1992,yoon10}.  Both single- and binary-star progenitors are
similarly theorized for the $\sim 1$\% of SNe Ibc associated with
long-duration gamma-ray bursts (GRBs), characterized by a central
engine-driven relativistic outflow (e.g., \citealt{mwh01,pods04}).  However, the critical difference
between ordinary SNe Ibc progenitors and those of GRB-SNe remains
debated (e.g.~\citealt{fmp+07}).

Recent studies of the host galaxies of SNe Ibc and GRB-SNe have
revealed surprising clues.  Specifically, explosion site metallicity
measurements suggest that GRB-SNe favor metal-poor environments more
often than SNe Ic \citep{mkk+08,emily10}.  In parallel efforts,
it has also been reported that SNe Ic reside in higher metallicity
environs than SNe Ib \citep{acj+10,mbf+10}.  These apparent
environmental differences have been interpreted as evidence for a
metallicity-dependent mass loss mechanism (e.g., \citealt{agk+10}),
such as line-driven winds \citep{Castor_et_al1975, Conti1978}.  
However, direct observational constraints
on the local (sub-parsec) circumstellar environments of the explosions
are required to test this hypothesis.

The mass loss histories of single- and binary-star progenitors each
produce unique density distributions in the circumstellar environment
(e.g.,~smooth stellar wind profile, dense common envelope shell). The dynamical
interaction between the expanding SN shockwave and the circumstellar
medium (CSM) gives rise to non-thermal radio and X-ray emission that,
in turn, traces the local mass distribution
\citep{c98}.  Radio and X-ray observations of young SNe Ibc are, in fact, the 
{\it only} way to reveal the mass loss histories of of their
progenitor stars in the final evolutionary stage.  Motivated thus,
the sample of SNe Ibc with detailed radio studies is growing,
primarily as part of the Very Large Array Intensive Study of Naked
Supernovae\footnote{Here we use ``naked'' to describe supernovae from
progenitor stars that have lost their hydrogen envelope prior to
explosion.}  (VISioNS;
\citealt{bkc02,bkf+03,Soderberg_et_al2005,Soderberg_et_al2006,skn+06,snb+06,Soderberg2007,Soderberg_et_al2008,Soderberg_et_al2010a, Soderberg_et_al2010b}).

The VISioNS study has revealed several key findings to date, including
(i) SNe Ibc radio luminosities span four orders of magnitude,
$L_{\nu}\approx 10^{25}-10^{29}~\rm erg~s^{-1}~Hz^{-1}$
\citep{Soderberg2007}, and (ii) nearly half of all radio-detected SNe
Ibc show evidence for small-scale (factor of $\sim 2$) light-curve
modulations (e.g.,~SN\,2003bg; \citealt{Soderberg_et_al2006}).  Both
of these findings are attributed to a broad diversity in the mass loss
histories of their progenitor stars \citep{Chevalier&Fransson2006}. In
parallel efforts, detailed radio studies of GRBs have
revealed similar diversity in their circumstellar environments (e.g., \citealt{clf04}),
however, a direct comparison of the two samples has yet to be
conducted.
  
Here we present extensive radio observations for three Type Ibc SNe --
2004cc, 2004dk, and 2004gq -- all of which show large (factor of $\sim
10$) flux modulations at late epochs.  Through our detailed modeling
of the radio light-curves, we derive the properties of the shockwave
(velocity, energy) and those of the circumstellar medium (density
profile, progenitor mass loss rate).  In all three SNe, we attribute
the observed flux modulations to density modulations in the local
environment. We compare the radio properties with those of other
SNe Ibc and nearby central engine-driven explosions, including GRB-SNe
1998bw \citep{Kulkarni_et_al1998} and 2006aj \citep{skn+06} 
and the relativistic SN\,2009bb \citep{Soderberg_et_al2010a}.  In
\S\ref{sec:VLA} we present the observations and in
\S\ref{sec:modeling} we describe the dynamical model used to extract
the properties of the shockwave and local environment and present the
results of modeling the early radio data. In
\S\ref{sec:bumps} we compare the models with the late-time radio data 
characterized by flux modulations.  We discuss the implications of the
observed flux variations in the context of circumstellar density
modulations in \S\ref{sec:discussion} and, finally, in light of
different channels for progenitor mass loss.

\section{Radio Observations}
\label{sec:VLA}

We observed SNe 2004cc, 2004dk, and 2004gq with the Very Large
Array\footnote{The National Radio
Astronomy Observatory is a facility of the National Science Foundation
operated under cooperative agreement by Associated Universities, Inc.}
(VLA) shortly after optical discovery as part of the VISioNS survey.
In each case we detected a radio source coincident with the optical
position and subsequently initiated an intense VLA follow-up campaign
to study the temporal and spectral evolution of the radio emission.
In Table~1 we summarize the basic properties of these SNe, including
the spectroscopic classification, host galaxy distance estimate, and
approximate explosion date. We adopt host galaxy distance estimates
and integrated apparent magnitudes from NED\footnote{\tt
http://nedwww.ipac.caltech.edu}, including cosmology independent
distances, when available.

Radio data were collected at multiple frequencies between 1.43 and
43.3 GHz (Tables~\ref{table:2004cc}-\ref{table:2004gq}).  The data
were obtained in various array configurations as denoted in the
Tables.  All VLA observations were taken in the standard continuum
observing mode with a bandwidth of $2\times 50$ MHz.  At observing
frequencies above 22 GHz, we included reference pointing scans to
correct for the systematic 10-20 asec pointing errors of the VLA
antennas.  We used primary calibrators 3C48, 3C147, and 3C286 for flux
calibration, while phase referencing was performed using a calibrator
within 10 degrees.  Data were reduced using standard packages within
the Astronomical Image Processing System (AIPS).  We fit a Gaussian
model to the radio SN in each observation to measure the integrated
flux density.

\begin{figure}
\plotone{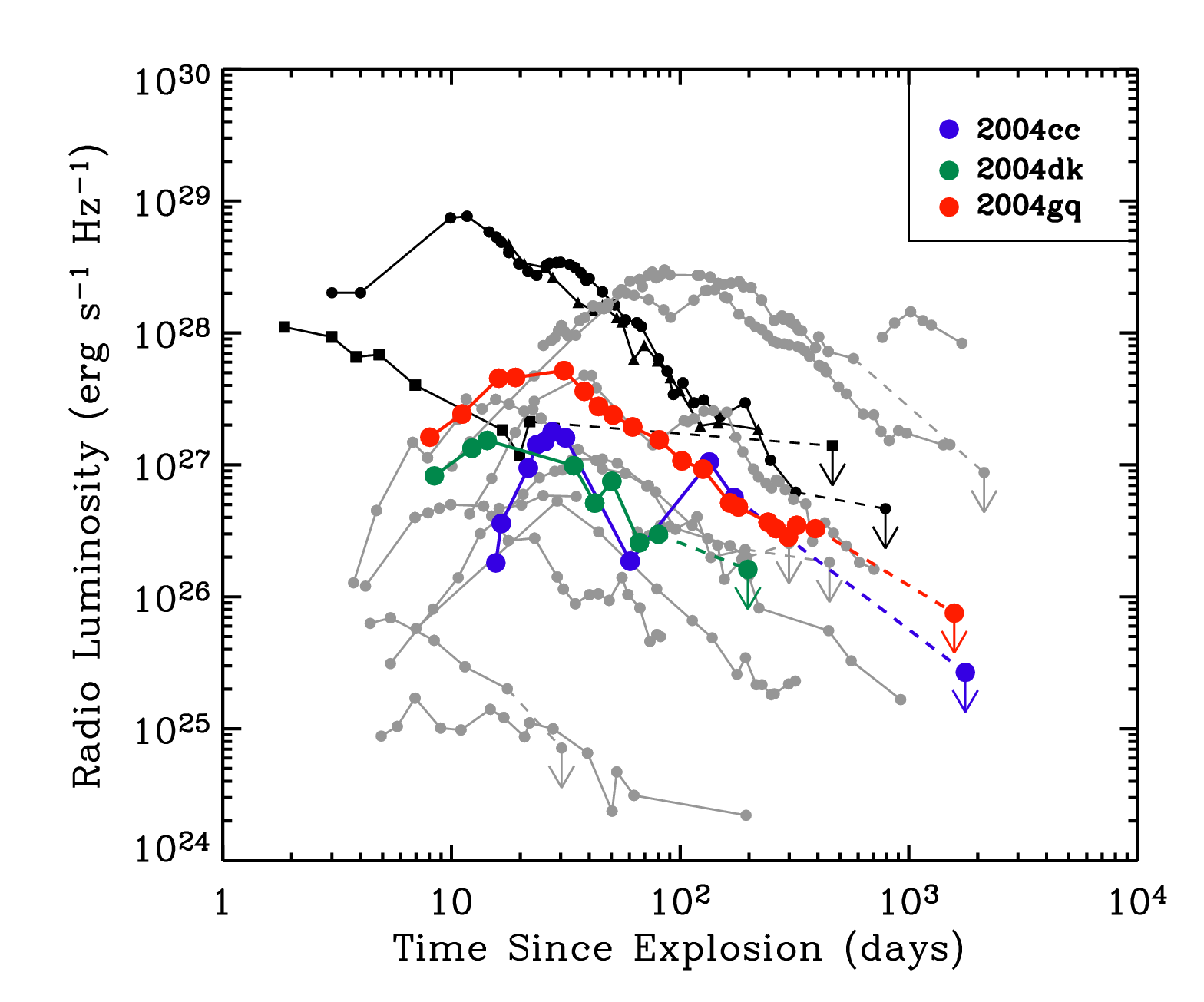}
\caption{Radio light curves for 2004cc (blue), 2004dk (green), 2004gq (red), 
are compared with other Type Ibc supernovae, including SNe 1983N
\citep{spw84}, 1990B \citep{vsw+93}, 1994I \citep{wei06}, 2001ig
\citep{Ryder_et_al2001}, 2002ap \citep{bkc02}, 2003L
\citep{Soderberg_et_al2005}, 2003bg \citep{Soderberg_et_al2006},
2007gr \citep{Soderberg_et_al2010b}, 2008D
\citep{Soderberg_et_al2008}, 2008ax \citep{rpb+09}, and 2011dh \citep{krauss11} all shown in grey.  The radio afterglows
of nearby engine-driven explosions within a similar volume are shown
in black, including GRB-SN\,1998bw (circles;
\citealt{Kulkarni_et_al1998}), XRF-SN\,2006aj (squares;
\citealt{skn+06}), and SN\,2009bb (triangles;
\citealt{Soderberg_et_al2010a}).  SNe 2004cc, 2004dk, and 2004gq are
similar in radio luminosity to ordinary radio SNe and less luminous
than engine-driven explosions on a similar timescale.}
\label{fig:lccompare}
\end{figure}

\begin{figure}
\plotone{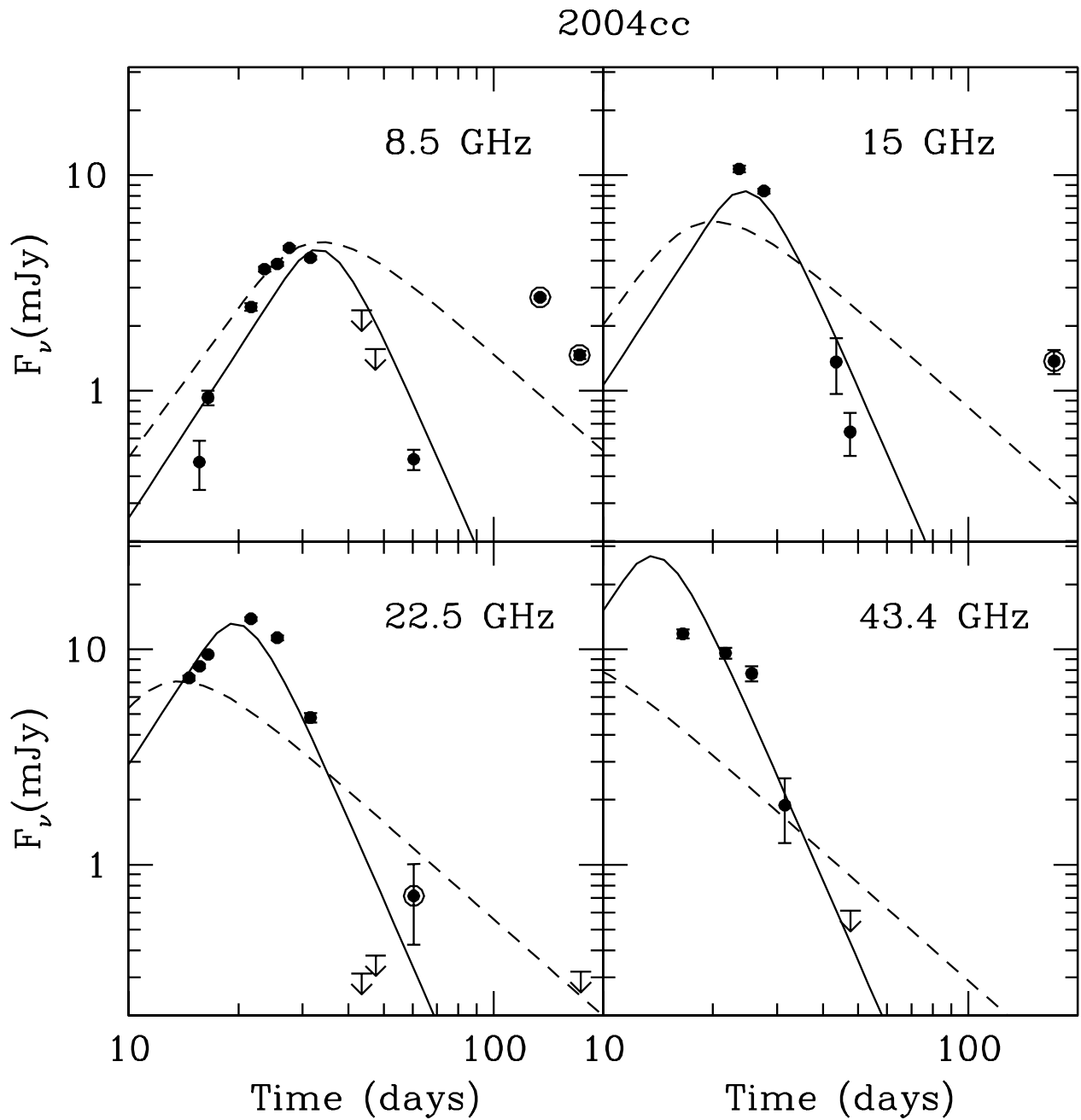}
\caption{Radio observations of SN\,2004cc are shown various frequencies and
compared to our dynamical models.  Dashed lines represent
the model parametrization assuming a density profile of $n_e\propto r^{-2}$.  
We find a significantly better fit (solid lines) associated with a 
steeper density profile.  Encircled data points indicate an
unusual late-time re-bightening and were not included in the model fits.}
\label{fig:lightfits_cc}
\end{figure}

\begin{figure}
\plotone{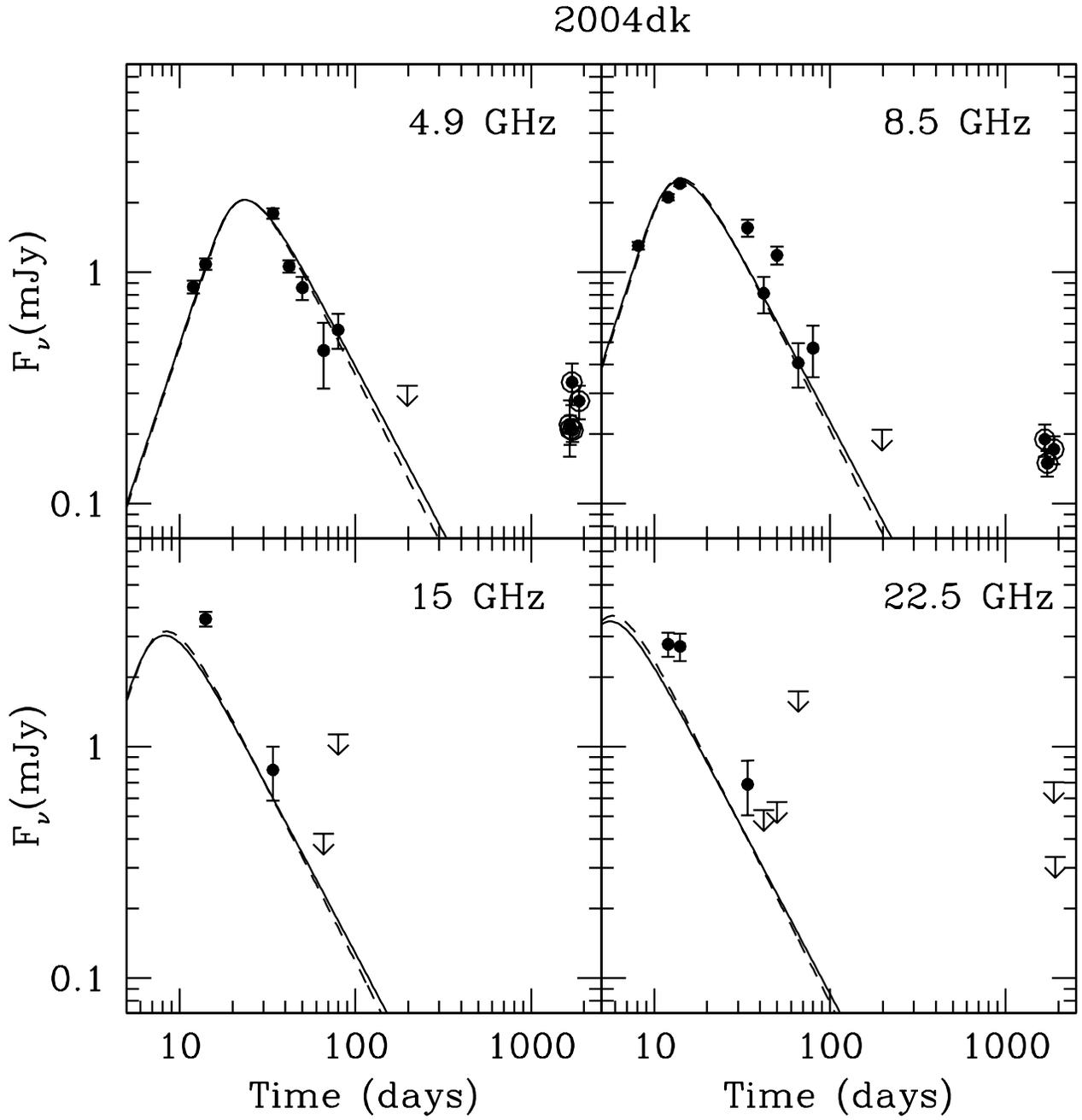}
\caption{Radio observations of SN\,2004dk are shown various frequencies and
compared to our dynamical models.  The solid lines represent the best
fit model parametrization, while the dashed lines represent a model in
which the density profile to fixed as $n_e\propto
r^{-2}$. Encircled data points at indicate an unusual late-time
re-bightening and were not included in the model fits, shown in detail
in Fig. 4.}
\label{fig:lightfits_dk}
\end{figure}

\begin{figure}
\plotone{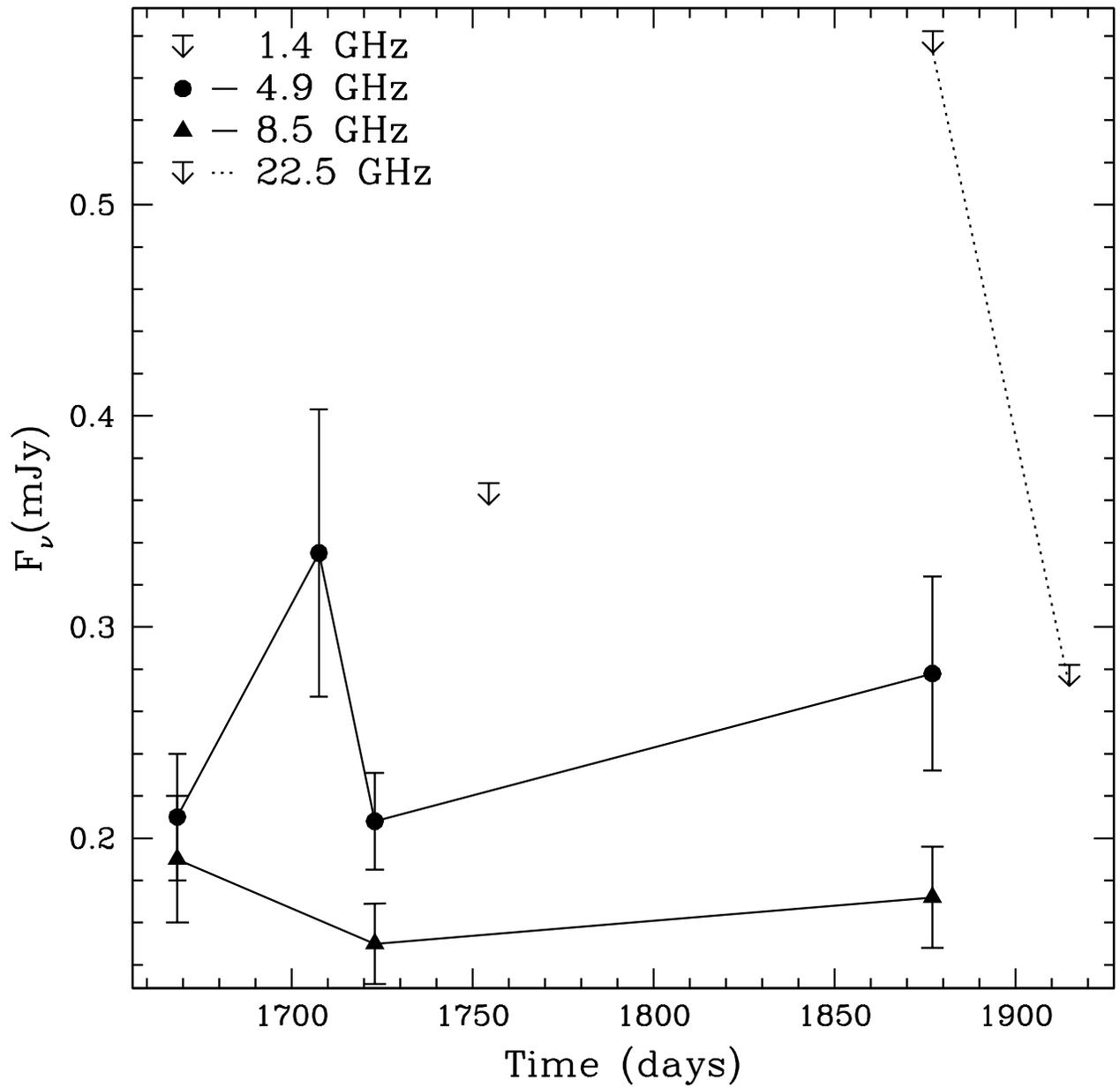}
\caption{Radio observations of SN\,2004dk at late time, during the
  rebrightening episode, are shown at several frequencies.  The SN is
optically thin throughout the rebrightening.}
\label{fig:late_time}
\end{figure}

\begin{figure}
\plotone{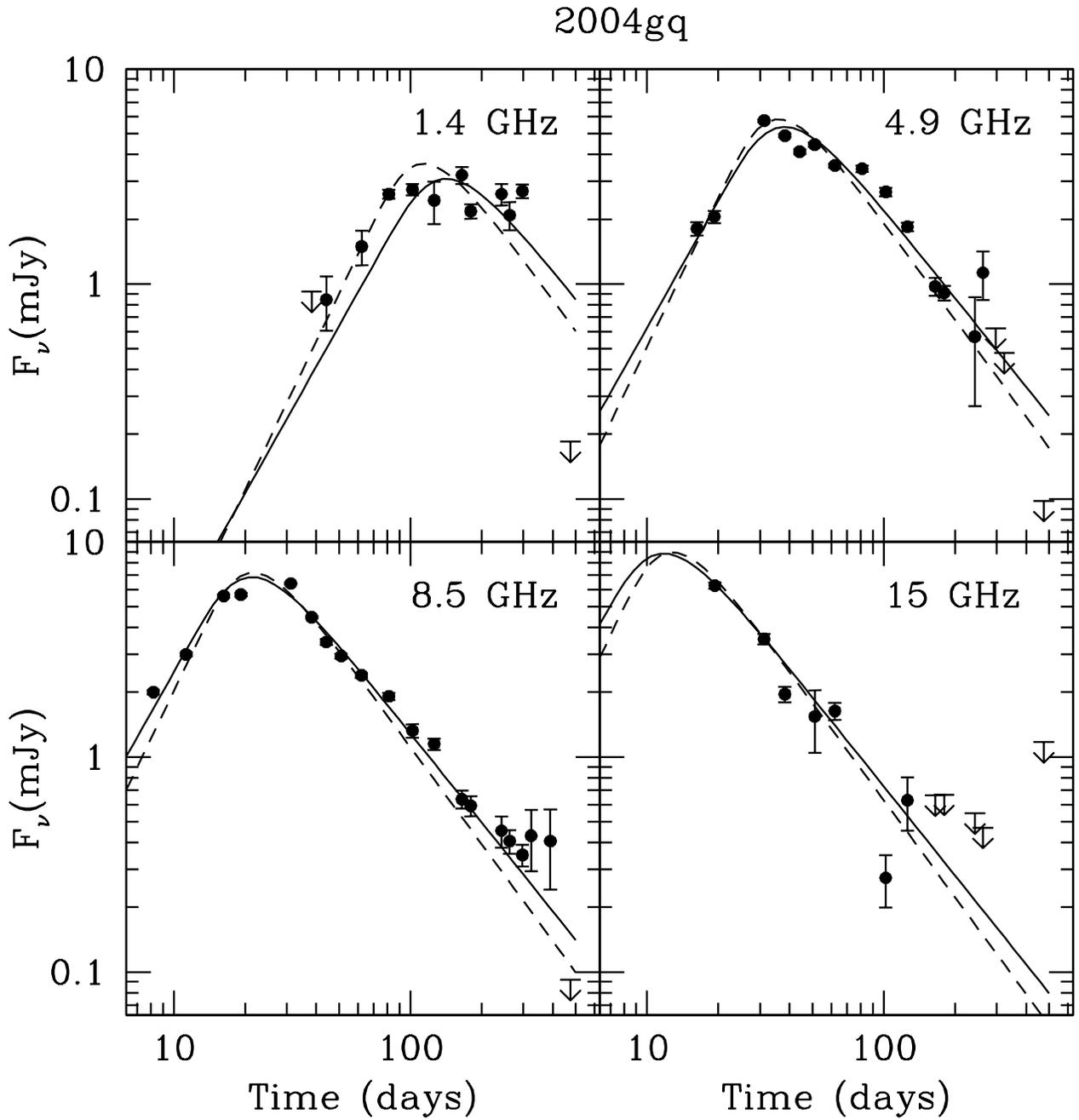}
\caption{Radio observations of SN\,2004dk are shown various frequencies and
compared to our dynamical models.  The solid lines represent the best
fit model parametrization, while the dashed lines represent a model
parametrization in which the density profile is fixed to $n_e\propto
r^{-2}$.  The deep limits at late-time indicate an unusual steepening of
the radio flux and were not included in the model fits.}
\label{fig:lightfits_gq}
\end{figure}

The radio light-curves for the three SNe are shown in
Figures~\ref{fig:lccompare}-\ref{fig:lightfits_gq}, spanning $\Delta t
\approx 8-1900$ days after explosion, and are discussed in detail
below.  Their peak spectral luminosities span $L_{\nu}\approx
(2-6)\times 10^{27}~\rm erg~s^{-1}~Hz^{-1}$, similar to other SNe Ibc
and a factor of $\sim 100$ below the radio afterglow luminosities of
nearby GRB-SNe and the engine-driven SN\,2009bb.  Intriguingly, all
three SNe reveal evidence for large-scale (factor of $\sim 10$)
light-curve variations at late time.  We estimate the ``abruptness''
of the flux variations by determining the time interval between the
epoch at which the modulation is first seen and the epoch of the
previous observation ($\delta t$) and normalize by
the time since explosion, $\Delta t$.

\subsection{SN 2004cc}
\label{ssec:obs2004cc}

SN\,2004cc was discovered on 2004 Jun 10.26 UT by the Lick Observatory
Supernova Search (LOSS; \citealt{li00}) using the Katzmann Automatic
Imaging Telescope (KAIT; \citealt{Monard&Li2004}).  The host galaxy,
NGC 4568, is classified as SAbc with an absolute $B-$band magnitude of
$M_B\approx -19.6$ mag assuming a distance of $d\approx 18$ Mpc.
Spectral analysis revealed SN\,2004cc to be a highly reddened SN Ic
\citep{Matheson_et_al2004,Foley04}.  An unfiltered light-curve indicates a
peak optical magnitude of $M\approx -16.2$ mag, less luminous than
typical SNe Ibc \citep{llc+10}.  However, this estimate does not
account for host galaxy extinction which is likely to be significant
(see \citealt{dsg+10} for a discussion of extinction diagnostics for
SNe Ibc).

Our first VLA observation was carried out on 2004 Jun 17 UT, roughly
$\Delta t\approx 15$ days after explosion.  We monitored the radio
emission from SN\,2004cc at frequencies, $\nu=8.5, 15.0, 22.5$ and
43.4 GHz through $\Delta t\approx 1700$ days (Table~\ref{table:2004cc}
and Figure~\ref{fig:lightfits_cc}).  At 8.5 GHz, the lightcurve
reaches a peak of $F_{\nu}\approx 4.5~\rm mJy$ at $\Delta t\approx 28$
days with a spectral luminosity of $1.7 \times 10^{27}$
$\rm{erg~s^{-1}~Hz^{-1}}$.  An abrupt steepening of the observed flux
was observed at all frequencies at $\Delta t\approx 40$ days after
explosion with a temporal profile at least as steep as $\Delta
F_{\nu}\propto t^{-3.4}$ ($\nu=8.46$ GHz). Subsequently, the SN
re-brightened at $\Delta t\approx 135$ days.  We estimate the
abruptness of these two episodes to be $(\delta t/\Delta t)\approx
0.27$ and $(\delta t/\Delta t)\lesssim 1.2$, respectively.

\subsection{SN 2004dk}
\label{ssec:obs2004dk}

SN\,2004dk was discovered by KAIT on 2004 Aug 1.19 UT
\citep{Graham&Li2004}.   The host galaxy, NGC 6118 (morphology SAcd), has an
absolute magnitude of $M_B\approx -19.4$ at $d\approx 23$ Mpc.
Although SN\,2004dk was originally classified as Type Ic, it was
re-classified as a Type Ib following the detection of helium features
\citep{Filippenko_et_al2004}.  A late-time spectrum
published by \citet{Maeda_et_al2008} reveals an ordinary hydrogen-poor SN Ib
spectrum.  Detailed optical photometry for SN\,2004dk were reported by
\citet{dsg+10} indicating a broad light-curve with a peak magnitude of
$M_{V}\approx -18.2$ mag.

We first observed the SN with the VLA on 2004 Aug 7 UT, roughly
$\Delta t\approx 8$ days after explosion.  Table~\ref{table:2004dk}
summarizes our observations of SN 2004dk, taken at frequencies of
$\nu=4.9,~8.5,~15.0$ and 22.5 GHz.  Comparable in spectral luminosity
to SN\,2004cc, it reaches a peak of of $L_{\nu}=1.6 \times 10^{27}$
$\rm{erg~s^{-1}~Hz^{-1}}$ at $\Delta t\approx 14$ days after explosion
in the $\nu=8.5$ GHz band.  Similar to the case of SN\,2004cc, radio
monitoring revealed a re-brightening at $\Delta t\approx 4$ years
after explosion \citep{shv+09} which was not evident in the previous
observation at $\Delta t\approx 200$ days.  We therefore estimate the
abruptness of the modulation to be $\lesssim 7.7$.

We note that X-ray emission was also detected from SN\,2004dk with XMM
on 2004 Aug 12 UT ($\Delta t\approx 13$ days) with a flux, $F_X=2.7\pm
0.8\times 10^{-14}~\rm erg~cm^{-2}~s^{-1}$ (0.5-8 keV;
\citealt{poo07,shv+09}).  At the distance of NGC\,6118, this implies
an X-ray luminosity of $L_X\approx 2\times 10^{39}~\rm erg~s^{-1}$ and
comparable to the X-ray luminosities of other ordinary SNe Ib observed
on a similar timescale (e.g., SN\,2008D;
\citealt{Soderberg_et_al2008}).  At this epoch, the spectral index
($\beta$; $F_{\nu}\propto
\nu^{\beta}$) observed between the radio and X-ray bands is $\beta_{RX}\approx
-0.77\pm 0.03$.

\subsection{SN 2004gq}
\label{ssec:obs2004gq}

SN\,2004gq was discovered independently by KAIT and the Stazione
Astronomica di Sozzago Supernova Search on 2004 Dec 11.36 UT
\citep{Pugh_et_al2004}.  The host galaxy, NGC 1832 (morphology SBbc) 
and has an absolute $B-$band magnitude of $M_B\approx -20.1$ mag at a
distance of $d\approx 26$ Mpc.  SN\,2004gq was also originally
classified as Type Ic, but was later re-classified as Type Ib after
the appearance of strong helium features in a spectrum obtained on
2005 Jan 7.3 UT
\citep{Modjaz&Falco2005}.  As in the case of SN\,2004dk, late-time
spectroscopy by \citet{Maeda_et_al2008} indicated an ordinary
hydrogen-poor SN Ib spectrum.  Detailed optical photometry for
SN\,2004gq was presented by
\citet{dsg+10} revealing a peak magnitude of $M_{V}\approx -17.6$ mag.

We initiated VLA observations of SN\,2004gq on 2004 Dec 16 UT, roughly
$\Delta t\approx 8$ days after explosion.  Our observations of
SN\,2004gq are summarized in Table~\ref{table:2004gq} spanning
frequencies, $\nu=1.4, 4.9, 8.5$ and 15 GHz.  At 8.5 GHz, the SN
reaches a peak luminosity of $L_{\nu}\approx 5.5 \times 10^{27}$
$\rm{erg~s^{-1}~Hz^{-1}}$ at $\Delta t\approx 21$ days.  At $\Delta
t\approx 400$ days, our observations reveal evidence for an abrupt
{\it fading} at all frequencies.  There are no subsequent radio
observations available to determine if the emission later
re-brightened.  We estimate the abruptness factor of the radio
steepening to be $0.21$ and the power law decay to be at least as
steep as $F_{\nu}\propto t^{-5.9}$.

\section{A Model for the Radio Emission}
\label{sec:modeling}

Supernova observations are typically limited to the optical and
near-IR bands where the signal is dominated by thermal emission from
the inner, slow-moving layers of ejecta that carry the bulk of the
kinetic energy.  On the other hand, radio observations trace the
outer, fastest-moving layer, the ``shockwave'', to which only
0.01-0.1\% of the total energy is coupled (e.g.,~\citealt{mm99}).
Long duration GRBs are the notable exception, for which the relativistic jets and
slower SN ejecta carry comparable energy, each approaching, and
sometimes exceeding, $E\sim 10^{51}$ erg (e.g.,~GRB\,030329;
\citealt{bkp+03,mdt+03,fsk+05}).

As the shockwave plows through the circumstellar medium, it
shock-accelerates electrons into a power-law distribution of
relativistic energies, $N(\gamma)\propto \gamma^{-p}$, above a minimum
Lorentz factor, $\gamma_m$.  The shockwave also generates amplified
magnetic fields which cause the electrons to gyrate, emitting
synchrotron radiation that peaks in the cm-band in the first days to
weeks following the SN explosion.  The synchrotron emission is
suppressed at low frequencies due to absorption processes.  In the
case of SNe Ibc, \citet{c98} showed that the dominant absorption
process is internal synchrotron self-absorption (SSA), whereas
external absorption processes (e.g. free-free) may contribute
significantly for some SNe II plowing into denser 
circumstellar environments with slower shock speeds (e.g.,~\citealt{fb98}).
Self-absorption defines the spectral peak frequency for the synchrotron
spectrum, $\nu_p$, below which the radio spectral index is $\beta=5/2$
(optically-thick) and above $\nu_p$ the spectral index is
$\beta=-(p-1)/2$ (optically-thin;
\citealt{Rybicki&Lightman1979}).  For SNe Ibc, a value of $p\approx 3$
is typical \citep{c98}.  As the shockwave expands, the optical depth
to SSA decreases, causing $\nu_p$ to cascade to lower frequencies with
time.

\subsection{Radio Spectral Energy Distributions}
\label{ssec:firstfit}

\begin{figure}
\plotone{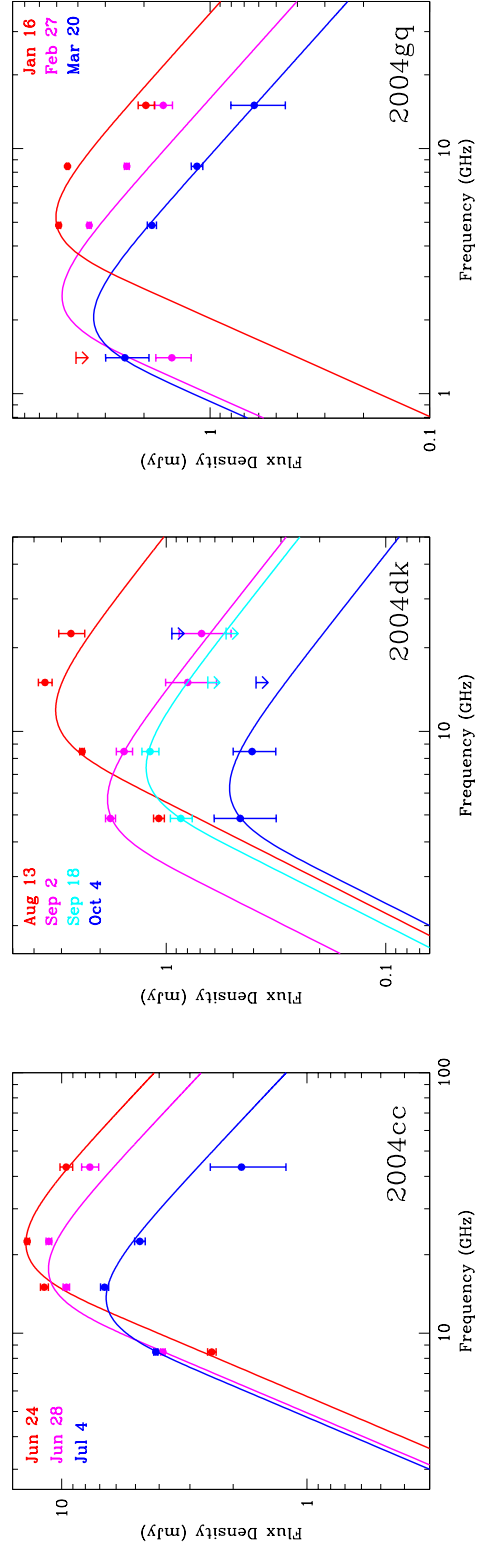}
\caption{Multi-frequency radio observations of SNe 2004cc, 2004dk, and 2004gq are well described by an SSA spectrum with $F_{\nu}\propto \nu^{5/2}$ in the optically thick regime and $F_{\nu}\propto \nu^{-1}$ in the optically thin regime,
associated with an electron index of $p\approx 3$.}
\label{fig:epfit}
\end{figure}

In our preliminary analysis of the SN data, we considered the
multi-frequency radio observations from each epoch to examine the
properties of the synchrotron spectrum at various stages in the radio
light-curve evolution.  As shown by \citet{c98}, a radio SN spectral energy
distribution characterized by SSA is well described as, $F_{\nu} = 1.582 F_{\nu_p}
({\nu}/{\nu_p})^{5/2} [1-{\rm exp}(-(\nu/\nu_p)^{-(p+4)/2})]$.  For each
radio spectrum, we used a $\chi^2$ minimization technique to determine
$p$, $\nu_p$, and the associated flux density at the spectral peak,
$F_{\nu,p}$. The fitting routine was carried out independently 
at each epoch for which we obtained observations in at least three
frequencies.  The radio emission from all three SNe is well
described by a SSA spectrum across multiple epochs (Figure~
\ref{fig:epfit}).  In each case, the electron 
energy index is consistent with $p\approx 3$ and the optically thick
spectral index is not steeper than $\beta\approx 5/2$ indicating that
any contribution from external absorption processes (e.g., free-free
absorption) is negligible.  As can be seen from the Figures,
$\nu_p$ cascades to lower frequencies with time and is accompanied by
a decrease in $F_{\nu,p}$.

We make the reasonable assumption that the post-shock energy density
is equally fractioned between relativistic electrons ($\epsilon_e$)
and amplified magnetic fields ($\epsilon_B$), i.e. equipartition, with
the remaining energy coupled to shocked protons
($\epsilon_p$).  In this framework, the radius of the emitting
material may be estimated from the spectral parameters measured at each epoch,
$R\approx 5.7\times 10^{15}~(L_{\nu,p}/10^{27}~\rm
erg~s^{-1}~Hz^{-1})^{9/19} (\nu_p/8.5~\rm GHz)^{-1}$ cm.  Here we have
assumed that the emitting region occupies half of the total volume
enclosed within a spherical shockwave.  Combined with an estimate of
the explosion date, the shockwave radius enables an estimate of the
time-averaged shock velocity, $\overline{v}$.

Similarly, the strength of the amplified magnetic field may be
estimated from the properties of the radio spectrum as $B\approx 0.93
~(L_{\nu,p}/10^{27}~\rm erg~s^{-1}~Hz^{-1})^{-2/19} (\nu_p/8.5~\rm
GHz)$ G.  From this estimate, the total internal energy density of the radio
emitting material is, $U\approx B^{~2}/8\pi \epsilon_B$.  It is
reasonable to assume that the partition fractions are each constrained
to be $(\epsilon_e,\epsilon_B)\lesssim 1/3$ \citep{c98}.  A further
constraint is implied by the shock velocity and the requirement that
$\gamma_m\gtrsim 1$ leads to a lower limit on the electron partition
fraction, $\epsilon_e\gtrsim 0.1(\overline{v}/0.2c)^{-2}$ for $p\approx 3$
\citep{Soderberg_et_al2005,Chevalier&Fransson2006,Soderberg_et_al2010b}.
In line with the typical velocities measured for SNe Ibc shockwaves,
$\overline{v}\approx 0.2c$, we make the reasonable assumption that
$\epsilon_e=\epsilon_B=0.1$ throughout this paper.

\subsection{A Dynamical Model}
\label{ssec:secondfit}

We adopt the formalism of \citet{Soderberg_et_al2005} in modeling the
dynamics of the shockwaves. In this model, the shockwave parameters
evolve self-similarly with time, $R\propto t^{\alpha_R}$ and $B\propto
t^{\alpha_B}$.  As shown by \citet{c82}, the expansion of the slightly
decelerated shockwave is characterized by $\alpha_R=(n-3)/(n-s)$ where
$n$ describes the density profile of the ejecta, $\rho_{\rm ej}\propto
r^{-n}$, and $s$ describes the density profile of the circumstellar
medium, $\rho_{\rm CSM}\propto r^{-s}$.  The self-similar model is
constrained such that $n\ge 5$ and in the case of a stellar wind
density profile, $s=2$ and $\alpha_B=-1$.  Using this general
dynamical prescription, we fit the flux density measurements at each
epoch and frequency simultaneously.

The flux density is parametrized as 
\begin{align}
\begin{split}
F_\nu(t,\nu) = & C_f \left(\frac{t}{t_0}\right)^{(4\alpha_R - \alpha_B)/2} [1 - e^{-\tau_\nu(t)}] \\
& \times \nu^{5/2} F_3(x) F_2^{-1}(x)
\end{split}
\end{align}
in erg/cm$^2$/s/Hz where $F_2$ and $F_3$ are Bessel
functions and $x\equiv 2/3(\nu/\nu_m)$ with $\nu_m$ defining the
critical synchrotron frequency, $\nu_m\equiv 3\gamma_m^2 q B/(4\pi m_e
c)$. The optical depth to SSA, $\tau_\nu$, evolves with time as
\begin{align}
\begin{split}
\tau_\nu(t,\nu) = & C_\tau \left(\frac{t}{t_0}\right)^{(3+p/2)\alpha_B + (2p-3)\alpha_R - 2(p-2)} \\
&\times \nu^{-(p+4)/2} F_2(x)
\end{split}
\end{align}
where $C_f$ and $C_\tau$ are normalization constants of the flux
density and optical depth, respectively, and $t_0$ is a reference
epoch. By fitting for $C_f$, $C_{\tau}$, $\alpha_R$, and $\alpha_B$,
we derived estimates for the physical parameters, $R_0$, $B_0$, and
$\gamma_{m,0}$ at time, $t_0$.

The number density of emitting electrons may be estimated from these
parameters as, $n_e\approx B^2/(4\pi \gamma_m m_e c^2)$ where we
maintain the assumption that $p\approx 3$ and $\epsilon_e=\epsilon_B$.
Thus, the number density of shocked electrons scales temporally as
$\alpha_{n_e}={2\alpha_B-\alpha_{\gamma}}$ and radially as
$s=-(2\alpha_B-\alpha_{\gamma})/\alpha_R$.  The density of unshocked
electrons in the circumstellar medium is related by the shock
compression factor, $\eta$, such that $n_{\rm CSM}=n_e/\eta$.  The
progenitor mass loss rate follows directly as $\dot{M}\approx 2\pi n_e
m_p R^2 v_w$ where we have assumed a compression factor of $\eta=4$
and a helium-rich stellar wind (nucleon-to-electron ratio of 2).
Here, $v_w$ is the velocity of the progenitor wind and we adopt
$v_w=10^3~\rm km~s^{-1}$ as measured for Galactic Wolf-Rayet stars
\citep{cgv04,cro07}.  The temporal evolution of the inferred mass loss
rate is, $\dot{M}\propto t^{2\alpha_B+2}$, and is constant for a
standard stellar wind profile with a steady wind velocity ($s=2$).

\subsection{Modeling Results}
\label{sec:results}

\begin{figure}
\plotone{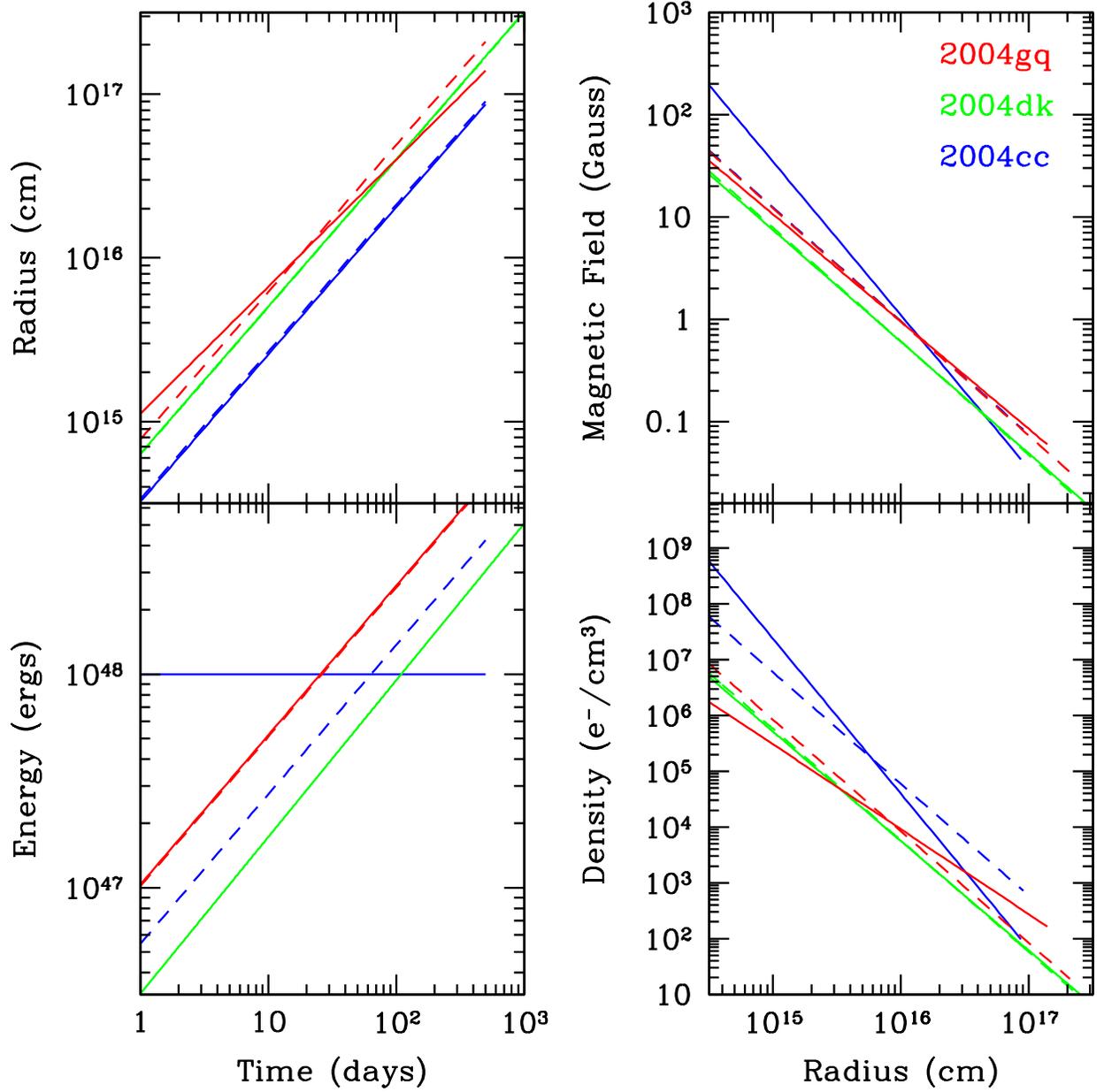}
\caption{The derived temporal and radial properties of the shockwave and local environment are compared for
SNe 2004cc (blue), 2004dk (green), and 2004gq (red).  The solid lines
represent the best-fit parametrization of our dynamical model while
the dashed lines correspond to a parameterization with a CSM denisty
profile fixed to $n\propto r^{-2}$.}
\label{fig:oneplot}
\end{figure}

For each SN, we model the early radio data only, excluding the unusual
late-time light-curve behavior in the fit.  The temporal and spectral
evolution of the SNe are well described by the dynamical model
(Figures~\ref{fig:lightfits_cc}-\ref{fig:lightfits_gq}).  For each SN,
we report the two model parametrizations which best represent the data
(see Table~\ref{tab:master}).  In the primary fit, we used a $\chi^2$
minimization technique to determine reasonable values for $C_f$,
$C_\tau$, $\alpha_R$, and $\alpha_B$.  In the ``standard'' model fit,
we fixed $s=2$, $\alpha_B=-1$ and $\alpha_R=0.9$ as expected in the
typical scenario where the forward shock is only slightly decelerated \citep{Chevalier&Fransson2006},
and fit only for $C_f$ and $C_{\tau}$.  We extracted estimates for the
physical parameters ($R$, $B$, $E$, and $n_e$) and their associated
temporal and radial evolution.  The results are detailed below for
each SN and summarized in Table~\ref{tab:master}.  We compare the
temporal and radial evolution of the physical parameters for these
three SNe in Figure~\ref{fig:oneplot}.

In the case of SN\,2004cc, the best fit is characterized by $\alpha_R\approx
0.9$, and a steep evolution of the magnetic field is required,
$\alpha_B\approx -1.4$.  Together these, imply a steep CSM density
gradient, $s\approx 2.8$.  At $t_0=10$ days, both models predict a
velocity of $\overline{v_0}\sim 0.1c$.  However, the models differ
regarding the energy, with the best fit model predicting an energy of
$E_0\approx 1.0\times 10^{48}$ erg while for the standard model,
$E_0\approx 2.7 \times 10^{47}$ erg.  Assuming the wind velocity
is unchanged, the inferred mass loss rate varies with time; we estimate $\dot{M_0}\approx 1.3\times
10^{-4}~\rm M_{\odot}~yr^{-1}$ 
at $t_0$ and just
prior to the late-time re-brightening episode it is a factor of $\sim
10$ lower, $\dot{M}\equiv 2.1\times 10^{-5}~\rm M_{\odot}~yr^{-1}$.

For SN\,2004dk we find that the standard model provides an excellent
description of the early radio data; our primary fit is not dissimilar
from the standard model.  We estimate a time-averaged
shockwave velocity of $\overline{v_0}\approx 0.2c$ and an energy of
$E_0\approx 1.7\times 10^{47}$ erg at $t_0=10$ days.  We find the
progenitor mass loss rate to be constant, $\dot{M_0}\approx 6.0\times 10^{-6}~\rm
M_{\odot}~yr^{-1}$.

In the case of 2004gq, we find that the best fit is nearly consistent
with a standard model, but with a shallower density profile, $s=1.5$.
In both models, the physical parameters at $t_0=10$ days are
essentially the same.  The implied time-averaged velocity is,
$\overline{v_0}\approx 0.25c$ and the energy is $E_0\approx 5\times
10^{47}$ erg.  The implied progenitor mass loss rate is
$\dot{M_0}\approx 9\times 10^{-6}~\rm M_{\odot}~yr^{-1}$ and is not
constant in the best fit model.  Just prior to the epoch of the
late-time flux steepening, the mass loss rate is $\dot{M}\approx
4\times 10^{-5}~\rm M_{\odot}~yr^{-1}$.

\section{Late-time Flux Variations}
\label{sec:bumps}

Intriguingly, the late-time data for all three SNe reveal unusual
behavior, in the form of an strong radio re-brightening for
SNe 2004cc and 2004dk, and an abrupt fading for SN\,2004gq.  These
late-time data were not included in the model fitting routines
described above.  For each SN, we measure the flux variation
by comparing the late-time measurements (or limits) to an
extrapolation of the model.  In Table~\ref{tab:master} we report
the late-time flux modulation factor for each SN.

Radio variability in young SNe is typically caused by modulations in
the circumstellar density profile, but engine-driven SNe may also
produce radio variability due to off-axis ejecta components (jets) or
prolonged central engine activity (see
\citealt{Soderberg_et_al2006} for a discussion of radio SN
variability).  In the case of CSM density modulations, the flux
variations are accompanied by a significant and abrupt shift in the
location of synchrotron self-absorption frequency.  This unique
observational clue has enabled circumstellar density variations to be
inferred for several core-collapse SNe exhibiting radio variability
\citep{Ryder_et_al2001,Soderberg_et_al2006}.

\citet{c98} show that for SNe dominated by SSA, 
$F_{\nu}\propto R^3 N_0 B^{(p+1)/2}$, where $N_0$ is the normalization
of the electron energy distribution at $\gamma_m\approx 1$. Assuming
$\epsilon_B$ is unchanged following an abrupt modulation in the
circumstellar density, we have $B^2\propto n_e v^2$ and thus expect
roughly $F_{\nu}\propto n_e^{(p+5)/4}$ or $F_{\nu}\propto n_e^2$ for
$p\approx 3$.  Doubling the CSM density therefore produces a factor of
$\sim 4$ increase in the optically-thin flux density.   Assuming that the wind
velocity is unchanged (a reasonable assumption since $v_w$ is roughly
set by the escape velocity) and the CSM density profile is
characterized by $s=2$, the implied mass loss rate modulation scales
similarly, $\Delta F_{\nu}\propto \Delta \dot{M}^2$.  In cases where
$s\neq 2$, we adopt $\Delta F_{\nu}\propto \Delta \langle
\dot{M}\rangle ^2$ where $\langle \dot{M}\rangle$ represents a 
radially-averaged mass loss rate.

\subsection{Mass Loss Variations}

SN\,2004cc shows evidence for a late-time re-brightening, nearly 100
times the flux density expected from an extrapolation of the model
fits to the early data (Figure~\ref{fig:lightfits_cc}).  The spectral
index between frequencies of $\nu=8.5$ and 15 GHz at $\Delta t\approx
130$ days is $\beta\approx -0.1$, indicating that the spectral peak
was between these two frequencies at that time.  However, $\nu_p$
cascaded through both frequencies several months earlier, indicating
that the re-brightening is associated with an increase of $\nu_p$.  We
therefore attribute the radio re-brightening to a significant
circumstellar density modulation.  A factor of $\sim 11$ increase in
the optically-thin flux corresponds to a CSM density jump by a factor
of $\sim 3.3$ and a mass loss rate of $\langle \dot{M}\rangle \approx 6.9
\times 10^{-5}$ $\rm{M}_\odot~yr^{-1}$.

More than four years after explosion, an abrupt re-brightening (factor
of $\sim 40$) is observed for SN\,2004dk and may be associated with a
flattening of the spectral index between $\nu=4.9$ and 8.5 GHz,
$\beta\approx -0.2$ (Fig. 4)\footnote{We note that the S/N of the
  late-time observations makes it difficult to precisely measure the
  spectral index.  Further observations of SN\,2004dk with the increased sensitivity
of the EVLA are planned to follow the long-term evolution of the radio
emission.}.  Similar to the case of SN\,2004cc, the
light-curves at these frequencies peaked much earlier and hence the
re-brightening is associated with an upward shift of $\nu_p$.  We
attribute this re-brightening to an increase in the circumstellar
density by a factor of $\sim 6$.  A density modulation of this size
implies an increased mass loss rate of $\langle \dot{M}\rangle \approx 3.5 \times
10^{-5}$ $\rm{M}_\odot~\rm yr^{-1}$.

In the case of SN\,2004gq, the late-time observations indicate a
dramatic decline in the light-curves at $\Delta t \approx 400$ days.
At this point, the flux density was below our detection threshold so a
spectral index measurement was not possible, however, we note that the
steepness of the decline ($F_{\nu}\propto t^{-6}$) is roughly
consistent with the evolution expected for spherical adiabatic expansion
losses of particles with an energy distribution of $p\approx 3$ and
flux freezing \citep{shk60,kp68}.  Attributing the flux decline to a
density drop, the measured upper limits imply a drop in the CSM
density by a factor of $\gtrsim 5$. We constrain mass loss rate
associated with this lower density wind to $\langle \dot{M}\rangle \lesssim 1.8
\times 10^{-5}$ $\rm{M}_{\odot}~yr^{-1}$.

In Figure~\ref{fig:oneplot}, we compare the circumstellar density
profiles for all three SNe as inferred from our dynamical modeling and
including the abrupt modulations.  Our radio observations point to
significant (factor of $\sim 3-6$) circumstellar density modulations
on radial scales of $r\approx (1-50)\times 10^{16}$ cm.  Assuming a
typical progenitor wind velocity of $v_w=10^3~\rm km~s^{-1}$, these
density modulations are traced back to mass loss variations spanning a
few years to a century prior to explosion.

\subsection{A Comparison to Other Radio SNe}

Our modeling of the early radio light-curves for SNe 2004cc, 2004dk,
and 2004gq indicate mass loss rates spanning $\dot{M}\approx
(0.6-13)\times 10^{-5}~\rm M_{\odot}~yr^{-1}$.  However, these estimates
vary by factors of $\sim 3-6$ following the observed late-time flux density
modulations.  In comparison with other radio SNe similarly observed on
sub-parsec scales, these SNe represent examples of the strongest flux
modulations ever reported.  To date, such large-scale (factor of $\sim
5-10$) radio modulations have only been reported for a few SNe,
including SN\,1987A
\citep{ball95,zan10}, SN\,1996cr \citep{bauer08}, SN\,2001em \citep{sks+05}, and
SN\,2007bg \citep{Soderberg_ATEL}.

Small-scale mass loss variations (factor of $\sim 2$) are more common
and have been inferred for several core-collapse SNe including 1979C
(Type IIL; \citealt{wvd+91}), 2001ig
(Type IIb; \citealt{Ryder_et_al2001}), 2003bg (broad-lined Type IIb;
\citealt{Soderberg_et_al2006}), and 2008ax (compact 
progenitor Type Ib/IIb; \citealt{rpb+09,cs10}). Radio variability was also
clearly seen for the relativistic central engine-powered 1998bw
(associated with GRB\,980425; \citealt{Kulkarni_et_al1998}) and 2009bb
\citep{Soderberg_et_al2010a,bsb+10}, both of which are
broad-lined SNe Ic.  However, it remains debated whether the
variability was produced by central engine activity, additional ejecta
components, or variations in the CSM density profile.  Therefore,
radio variability extends across the various SN Ibc sub-classes,
including helium-rich, helium-poor, broad-lined, engine-driven and
even hydrogen-rich (Type cIIb; \citealt{cs10}).  We stress, however,
that density-induced flux variations are not ubiquitous as evidenced
by several well-studied SNe with smooth radio light-curves including
SN\,1993J (IIb; \citep{wwp+07}), SN\,2003L (Ic;
\citealt{Soderberg_et_al2005}), SN\,2008D (Ib;
\citealt{Soderberg_et_al2008}), and GRB-SN\,2003dh (i.e.\,~the radio
afterglow of GRB\,030329; \citealt{bkp+03}), each observed on similar
sub-parsec radial scales.

We compare the radio-derived mass loss
rates for SNe 2004cc, 2004dk, and 2004gq with those of other local SNe
Ibc and engine-driven (GRB-SN) explosions within a comparable volume,
$d\lesssim 150$ Mpc.  The comparison sample of SNe represents a
literature compilation (see Figure~\ref{fig:lccompare} caption), with inferred
mass loss rates spanning $\dot{M}\approx 10^{-7}-10^{-2}~\rm
M_{\odot}~yr^{-1}$ and a distribution peak near $\dot{M}\approx
10^{-5}~\rm M_{\odot}~yr^{-1}$ (see also \citealt{Chevalier&Fransson2006}).  We derive mass loss rates
for SNe 2004cc, 2004dk, and 2004gq that overlap with the higher end of
this SN Ibc distribution.  In contrast, the engine-driven explosions
considered here populate the low end of the mass loss rate distribution,
$\dot{M}\approx (1-50)\times 10^{-7}~\rm M_{\odot}~yr^{-1}$.  

A modest dispersion in the $\dot{M}$ distribution may be expected due
to variations in the shock partition fractions ($\epsilon_e$,
$\epsilon_B$), however, these are unlikely to deviate strongly (e.g., 
$\gtrsim 100$) from equipartition (see \citet{Chevalier&Fransson2006} for 
discussion).  Indeed, in 
the case of SN\,2011dh the partition fractions were found to have a ratio 
$\epsilon_e/\epsilon_B\sim 30$ \citep{smz+12}.  For deviations from 
equipartition, the mass loss rates reported here would scale as $\dot{M}\propto 
\epsilon_e^{-8/19} \epsilon_B^{-11/19}$.  We emphasize, however, that this only 
{\it increases} the inferred mass loss rates.

Some
dispersion in the $\dot{M}$ distribution is also expected from an
intrinsic spread in progenitor wind speeds; we emphasize that the radio
observations only constrain the quantity, $(\dot{M}/v_w)$, and we have
assumed a constant wind, $v_w=10^3~\rm km~s^{-1}$ for each SN.  However, $v_w$ is
regulated by the progenitor escape speed and is not seen to vary by
more than a factor of a few in the set of well-studied Galactic WRs
(see \S\ref{sec:discussion}).  Therefore, the factor of $10^5$ spread
in the inferred late mass loss rates (last 10-100 years) of SN Ibc progenitors points to an
intrinsically broad dispersion\footnote{We note that the subset of
rare, radio luminous SNe are likely over-represented in this
literature compilation since they are preferentially detected (e.g.,
SN\,2003L;
\citealt{Soderberg_et_al2005,Chevalier&Fransson2006}).  A distance-limited study of radio SN 
luminosities, and in turn, progenitor mass loss rates is required to
determine the intrinsic dispersion in $\dot{M}$.  This is the focus of a
separate paper.}.  

In comparison to other supernova sub-types, the mass loss rates for these SN
Ibc progenitors are lower than those inferred for most Type IIn SNe,
revealing intense progenitor mass loss episodes with
$\dot{M}\approx 0.01-1~M_{\odot}~\rm yr^{-1}$ and wind velocities of
$v_w\approx 300~\rm km~s^{-1}$ in the decades leading up to explosion
\citep{smi08,kga+10}. The local CSM densities of SNe IIn are so
high that the radio emission typically becomes optically-thin only {\it
years} after the explosion in the cm-band such that the early radio
signal is suppressed below detectable limits (e.g., SN\,1988Z;
\citealt{wpv+02}).


\section{The Nature of CSM Density Modulations}
\label{sec:discussion}

While it is generally understood that the massive progenitor stars of
SNe Ibc (including relativistic SNe) lose their envelopes prior to
explosion, the mechanism(s) by which the material is removed and the
relevant timescale(s) remains debated \citep{slf+10}.  Here we review the
possible causes of circumstellar density modulations in core-collapse SNe.

Isolated Wolf-Rayet stars, the favored progenitors of SNe Ibc, are
seen to lose mass at a typical rate of $\dot{M}\approx (0.5-2)\times
10^{-5}~\rm M_{\odot}~yr^{-1}$ (including Galactic WN, WC, and WO subtypes;
\citealt{cgv04,cro07}) through line-driven winds traveling at $v_w\approx (0.7-2.5)\times 10^{3}~\rm km~s^{-1}$, propelled by radiation pressure and momentum coupling \citep{Castor_et_al1975,
Conti1978}.  These winds shape the immediate circumstellar environment
on sub-parsec scales.  The strength of such line-driven winds scales
almost linearly with metallicity ($\dot{M}\propto Z^{0.8}$;
\citealt{vd05}), and an upper limit on this process is set by
saturation of the metal lines at $\dot{M}\approx
10^{-4}~\rm{M}_{\odot}~yr^{-1}$. Small-scale clumping within the winds
due to turbulence can produce moderate density modulations (factor of
$2-4$; see \citealt{mof08}).  At similar circumstellar radii, modest
density modulations may be present due to subtle variations in the
stellar wind velocity.  The collision of winds in a tight binary
systems can also give rise to CSM density modulations in the immediate
vicinity of the progenitor.

At larger circumstellar radii (a few parsec) an abrupt density jump is
likely produced by a wind termination shock between the fast
Wolf-Rayet wind and the slower red supergiant (RSG) wind
\citep{glm+96}.  Numerical simulations indicate that the
structure of the wind-termination shock is shell-like, and
that the interaction of the SN shockwave with this density feature
gives rise to a strong radio and X-ray signal, accompanied by 
hydrogen recombination lines in the optical band (e.g.,~\citealt{ddb10}).

Massive circumstellar shells are also produced in binary progenitor
systems where a common envelope is ejected by the motion of a
companion star \citep{Podsi_et_al1992}. The envelope material moves more
slowly ($v_w\approx 10~\rm km~s^{-1}$) than the Wolf-Rayet wind and so
becomes accelerated and compressed into a thin dense shell.  The
interaction of the shockwave with the common envelope shell produces
an abrupt increase in the synchrotron emission and Balmer series
recombination lines visible in optical spectra.  Late-time radio and
optical observations of SN\,2001em revealed these characteristics
\citep{sgk04,sks+05}, motivating \citet{Chugai&Chevalier2006} to
suggest that a common envelope was lost at a rate of (2-10) $\times
10^{-3}$ $\rm{M}_\odot~yr^{-1}$ and located at $r\approx 7
\times 10^{16}$ cm at the time of the explosion.

Large density modulations may also be fueled by violent stellar
outbursts such as those observed from Luminous Blue Variable stars
(LBVs; \citealt{hd94}).  LBVs are thought to represent a short-lived
phase in the evolution of some supergiants, and immediately precede
the development of classic WR features \citep{cro07}.  Along this
line, LBVs and WRs both show evidence for strong stellar winds and
H-poor envelopes.  LBVs give rise to giant mass loss ejections
exceeding the saturation limit associated with line-driven winds and
are therefore attributed to metallicity-independent continuum-driven
winds and/or hydrodynamic eruptions \citep{Smith&Owocki2006}.  The
outbursts result in dense and massive circumstellar shells that expand
into a strong (and quasi-steady) wind.  Multi-wavelength emission is
observed from the resulting nebulae surrounding local LBVs on radial
scales of of $r\lesssim 1~\rm pc$ (e.g.,~\citealt{ubt+10}).

The long-lived and luminous multi-wavelength emission from Type IIn
SNe is commonly attributed to the abrupt interaction of the shockwave
with LBV-like mass ejections in the decades leading up
to the explosion (e.g.,~\citealt{cbc+04}).  A recent (and extreme) example is
the Type IIn SN\,2006gy, for which a progenitor mass loss rate of
$\dot{M}\approx 0.5~M_{\odot}$ has been inferred
(\citealt{ock+07,slf+07}; see
\citealt{vso+10} for a review).  An intriguing aspect of the SN IIn
mass loss histories is the synchronization required between the
violent ejection episodes and the ultimate explosion (see
\citealt{smi08} for a discussion).  An LBV-like eruption seen from the
Type Ib SN\,2006jc progenitor star two years prior to explosion
similarly indicates a synchronization, and in turn, a direct link
between SN Ibc and IIn progenitors
\citep{fsg+07,psm+07}.  At longer wavelengths, a link was also suggested by
attributing the episodic radio variability observed for Type Ib/IIb
SNe\,2001ig and 2003bg to S-Doradus pulsations of the progenitor star
(\citealt{kv06}, but see \citealt{Soderberg_et_al2006} and \citealt{cs10} for
a different interpretation).

\subsection{SNe 2004cc, 2004dk, 2004gq}

We consider the observed multi-wavelength properties for SNe
2004cc, 2004dk, and 2004gq in light of the various causes of
circumstellar density modulations discussed above.  As discussed in
\S\ref{sec:bumps}, the late-time radio data for these three SNe imply
mass loss rates spanning $\dot{M}\approx (2-7)\times 10^{-5}~\rm
M_{\odot}~yr^{-1}$ at circumstellar radii of $r\approx (1-50)\times
10^{16}$ cm.  Their late-time radio variability implies circumstellar
density modulations by a factor of $\sim 3-6$.  Moreover, early
optical photometry and late-time (nebular) spectroscopy are consistent
with observations of typical SNe Ibc and do not reveal any evidence
for unusually strong circumstellar interaction (e.g.,~H recombination
lines; \S\ref{sec:VLA}).

The intermediate-level density modulation factors and ordinary optical
properties observed for these SNe indicate that it is unlikely that their
shockwaves are interacting with a H-rich common envelope.  Attributing the abrupt
density modulations to a clumpy wind would imply that the clumps are
exceedingly large (comparable to the size of the expanding shockwave)
in order to produce an observable upward shift in the self-absorption
frequency.  This is inconsistent with the basic picture for
small-scale, turbulence-driven WR wind clumps (e.g.\,
\citealt{mof08}).  Interaction with the wind termination shock should
occur at significantly larger radii (several pc) so we find this
scenario unlikely.  However, it is possible to reduce the radius of
the termination shock if these SNe are embedded in highly
pressurized environments \citep{clf04} or if their progenitors were 
runaway stars prior to explosion \citep{cyl+07, vla+08}.

\begin{figure}
\plotone{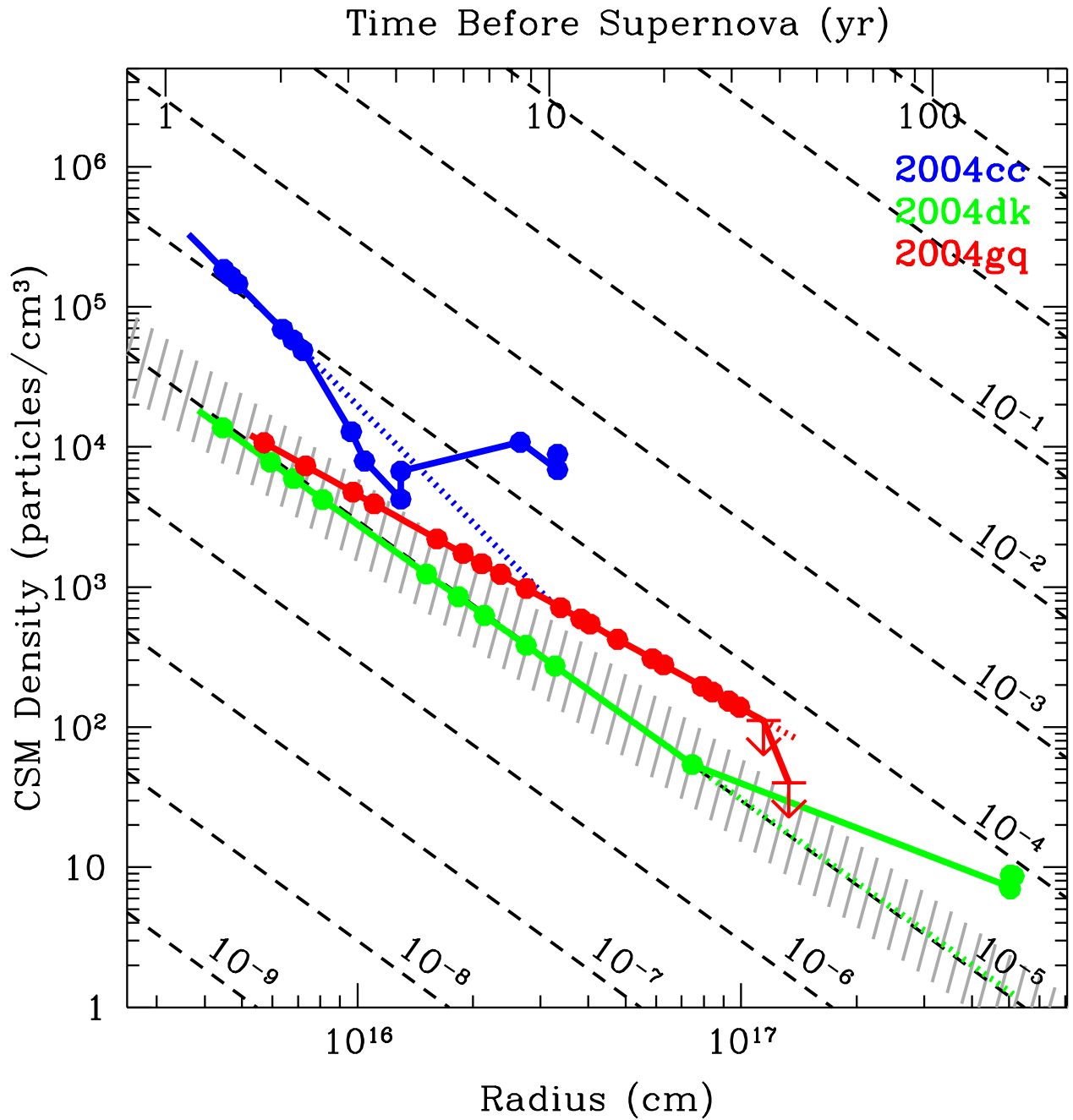}
\caption{The density profile of shocked particles is shown as a function of 
circumstellar radius for SNe 2004cc (blue), 2004dk (green), and 2004gq
(red).   Modeling of the early data imply density values shown by the
color dotted lines.  The abrupt deviations from these fits at later epochs are
associated with CSM density modulations. Radially averaged mass
loss rates are shown for comparison (black dashed lines).  The grey
hatched region marks the
observed range of mass loss rates for Galactic Wolf-Rayet stars.
Assuming a wind velocity of 1000 km/s, these density modulations could
be attributed to variable progenitor mass loss on a timescale
of $\sim 10-100$ years before explosion (top axis).}
\label{fig:density}
\end{figure}

In comparison to mass loss rates measured for Galactic Wolf-Rayet
stars, the inferred circumstellar densities inferred for these three
SNe are at the high end of the distribution
(Figure~\ref{fig:density}).  They show overlap with the mass loss
rates inferred for radio luminous SNe 2003L and 2003bg on similar
radial scales ($\dot{M}\approx 1-3\times 10^{-4}~\rm
M_{\odot}~yr^{-1}$; \citealt{Chevalier&Fransson2006}), which are
roughly consistent with the saturation limit for line-driven winds.

This, together with the abruptness of the mass loss variations and
their apparent synchronization with respect to the explosion date,
resembles the properties of SNe IIn.  Their unusual mass loss
properties may indicate some contribution from metallicity-independent
envelope-stripping mechanisms (e.g., continuum-driven winds and/or
hydrodynamic eruptions).  While the density enhancements inferred here
are not as extreme as those seen for Galactic LBVs or inferred for SN
IIn progenitors, their abrupt nature is reminiscent.  We note that recent 
theoretical efforts suggest that such abrupt mass loss episodes could be 
driven by pulsations or gravity waves induced by convective motions 
\citep{y+c10,q+s12}.

\subsection{A Metallicity Dependence?}

A key question raised by these observations is whether the
circumstellar environments inferred for some SNe are shaped by
metallicity-dependent line-driven winds or ``LBV-like''
metallicity-independent mass loss mechanisms.  While it is not
possible to measure the metallicity of the progenitor star
post-explosion, the metallicity of the explosion site on a sub-kpc
scale may serve as a reasonable proxy for that of the short-lived
progenitor system.  A comparison between the explosion site
metallicities and the radio-derived mass loss rates tests the
metallicity-dependence of the envelope-stripping mechanism.

As a rough estimate of the explosion site metallicity, we adopt the
absolute $B-$band magnitudes for the three host galaxies (Table~1) and
the luminosity-metallicity relation of \citet{tre04}.  We find host
galaxy metallicities of [12+log(O/H)]$\approx 8.9$, 8.8, and 9.0 for
SNe 2004cc, 2004dk, and 2004gq, respectively.  These estimates are
somewhat higher than the solar metallicity, [12+log(O/H)]$\approx
8.69$ \citep{ala+01}.  However, these values are {\it global} metallicity
estimates, and there is evidence that the explosion site metallicities
of SNe Ibc can be significantly different than those estimated
globally due to metallicity gradients within the host galaxy (see
\citealt{acj+10,mbf+10} for a discussion).  Along this line,
\citet{acj+10} recently reported an explosion site metallicity for
SN\,2004gq of [12+log(O/H)]$\approx 8.8$, lower than that estimated
globally for the host galaxy\footnote{We note, however, that the explosion site
metallicity was estimated using the emission line metallicity
diagnostic of \citet{pp04} which prevents a direct comparison of the
two $Z-$estimates.}.  Explosion site metallicities are currently
unavailable for SN\,2004cc and SN\,2004dk; however,
\citet{mbf+10} estimate that the offset between global and sub-kpc
metallicities for SNe Ibc is typically of order $\pm 0.2$ dex.  We
conclude that the explosion sites metallicities for these three SNe
are likely typical, probably not far from the Solar value and
consistent with the range of metallicities at which line-driven winds
operate efficiently.
 
Finally we note that the current range of explosion site metallicities
spans just 0.5 dex for ordinary SNe Ibc and engine-driven
explosions \citep{mbf+10}.  However, the
dispersion in radio-derived mass loss rates spans several orders
of magnitude, significantly broader than expected if the
$\dot{M}-Z$ relation is roughly linear.  This preliminary comparison
suggests that metallicity-dependent mass loss processes (e.g.,~line
driven-winds) are not the only mechanism by which the progenitor
envelope is removed in SNe Ibc and/or engine-driven explosions.

\section{Conclusions}
\label{sec:conclusion}

We present extensive radio observations and detailed modeling for the
three nearby Type Ibc SNe 2004cc, 2004dk, and 2004gq.  We show that
the physical properties of their shockwaves are typical of other SNe
Ibc while their circumstellar environments are characterized by
unusually strong density modulations.  We conclude with the following
points:

\begin{itemize}
\item{The radio spectra for all three SNe are well described by a  self-absorbed synchrotron spectrum with a relativistic $e^{-}$ population characterized by  $\gamma_m\sim 1-2$ and $p\approx 3$}. 
\item{Detailed modeling of the early radio data shows them to be consistent with free-expansion model in which a non-relativistic shock interacts with a stellar wind environment.  We extract shockwave velocities of $v \sim 0.1-0.25c$ and energies of $E \approx (2-10) \times 10^{47}$ erg at $\Delta t\approx 10$ days.}
\item{Each SN is characterized by a late-time radio re-brightening or
    fading which we attribute to abrupt and significant (factor of
    $\sim 3-6$) density modulations at circumstellar radii of,
    $r\approx (1-50)\times 10^{16}$ cm}.  The inferred mass loss rates
  rates lie at the high end of the  $\dot{M}$-distribution measured for Galactic WRs
\item{If we attribute the density modulations to variations in the
    progenitor mass loss rates then they approach the
    saturation limit for line-driven winds.   If the density modulations
are instead due to the collision with a wind termination shock, the
SNe must reside in highly pressurized environments {\it or} be runaways in order to squeeze the termination shock
to small circumstellar radii.} 
\item{The possible synchronization of the mass loss variations with the SN explosion 
is reminiscent of SN IIn progenitors.}
\item{A comparison of the radio-derived 
$\dot{M}$ estimates and explosion site metallicities for stripped envelope
SNe could shed light on the primary mass loss mechanism(s).}
\end{itemize}

Finally, we note that that the strong flux density
variations reported here are unlikely to be ubiquitous since they are
recoverable by radio campaigns designed to search for off-axis GRB
jets associated with SNe Ibc on timescales of months to years after explosion
\citep{pl98,pac01}.  Such radio searches
have uncovered strong late-time variability in less than $\sim 10\%$
of all SNe Ibc
\citep{bkf+03,snb+06}.  Future radio efforts 
will shed light on the variability of radio SNe over various time
baselines and therefore able to probe density modulations at a
wide-range of circumstellar radii. In particular, the improved
sensitivity of the Expanded Very Large Array \citep{pnj+09} will
enable detailed monitoring of radio SNe to longer timescales, enabling
the search for variable progenitor mass loss histories at larger
circumstellar radii, possibly associated with earlier pre-explosion epochs.

\smallskip

We thank our anonymous referee, Norbert Langer, Stan Owocki, Nathan Smith, Vikram
Dwarkadas, and Shri Kulkarni for helpful discussions.
S.W. and A.M.S. acknowledge support by the National Science Foundation
Research Experiences for Undergraduates (REU) and Department of
Defense Awards to Stimulate and Support Undergraduate Research
Experiences (ASSURE) programs under Grant no. 0754568 and by the
Smithsonian Institution.  

\bibliographystyle{apj1b}

\clearpage

\clearpage

\clearpage

\clearpage

\clearpage

\clearpage

\clearpage

\clearpage

\begin{deluxetable}{ccccccccc}
\normalsize
\tablecaption{SN Sample Properties}
\tablehead{
\colhead{Name} & \colhead{Host Galaxy} & \colhead{Distance\tablenotemark{1}} & \colhead{Host Magnitude} & \colhead{Disc. Date\tablenotemark{2}} & \colhead{Explosion Date} & \colhead{Type} & \colhead{$F_{p,8.5~\rm GHz}$} & \colhead{$t_{p,8.5~\rm GHz}$}  \\
\colhead{} & \colhead {} & \colhead{(Mpc)} & \colhead{($M_B$)} & \colhead{(UT)} & \colhead{(UT)} & \colhead{} & \colhead{(mJy)} & \colhead{(days)} 
}
\startdata
2004cc & NGC 4568 & $18\pm 2$ & $-19.6\pm 0.3$ & 2004 Jun 10 & 2004 Jun 2 & Ic & 4.5 & 32 \\
2004dk & NGC 6118 & $23\pm 2$ & $-19.4\pm 0.3$ & 2004 Aug 1 & 2004 Jul 30 & Ib & 2.5 & 14 \\
2004gq & NGC 1832 & $26\pm 5$ & $-20.1\pm 0.7$ & 2004 Dec 11 & 2004 Dec 8 & Ib & 6.9 & 21 \\ 
\enddata
\tablenotetext{1}{We adopt cosmology 
independent distances from NED-1D when available, otherwise we adopt the model-derived NED distances assuming $H_0=73~\rm km~s^{-1}~Mpc^{-1}$, $\Omega_{M}=0.27$, and $\Omega_{\Lambda}=0.73$.}
\tablenotetext{2}{We estimate the explosion date by the linear mean of the UT dates bridging the optical SN discovery and the most recent pre-discovery non-detection.}
\label{table:sne}
\end{deluxetable}

\newpage

\begin{deluxetable}{crrrrc}
\tablecaption{Observations of SN 2004cc}
\tablehead{
\colhead{Date} & \colhead{$F_{\nu,8.5~\rm GHz}$} & \colhead{$F_{\nu,15~\rm GHz}$} & \colhead{$F_{\nu,22.5~\rm GHz}$} & \colhead{$F_{\nu,43.4~\rm GHz}$} & \colhead{VLA} \\
\colhead{(UT)} & \colhead{(mJy)} & \colhead{(mJy)} & \colhead{(mJy)} & \colhead{(mJy)} & \colhead{Config.}
}
\startdata
2004 Jun 17.2 & ... & ... & 7.35 $\pm$ 0.18 & ... & D \\
2004 Jun 18.1 & 0.47 $\pm$ 0.12 & ... & 8.33 $\pm$ 0.16 & ... & D \\
2004 Jun 19.0 & 0.93 $\pm$ 0.07 & ... & 9.44 $\pm$ 0.11 & 11.79 $\pm$ 0.55 & D \\
2004 Jun 24.2 & 2.44 $\pm$ 0.10 & ... & 13.83 $\pm$ 0.29 & 9.60 $\pm$ 0.57 & D \\
2004 Jun 26.1 & 3.66 $\pm$ 0.10 & 10.69 $\pm$ 0.37 & ... & ... & D \\
2004 Jun 28.1 & 3.87 $\pm$ 0.09 & ... & 11.29 $\pm$ 0.30 & 7.70 $\pm$ 0.61 & D \\
2004 Jun 30.1 & 4.60 $\pm$ 0.09 & 8.44 $\pm$ 0.21 & ... & ... & D \\
2004 Jul 4.0 & 4.13 $\pm$ 0.08 & ... & 4.81 $\pm$ 0.24 & 1.89 $\pm$ 0.63 & D \\
2004 Jul 16.0 & < 2.36 & 1.36 $\pm$ 0.40 & < 0.31 & ... & D \\
2004 Jul 20.0 & < 1.56 & 0.64 $\pm$ 0.15 & < 0.38 & < 0.61 & D \\
2004 Aug 1.9 & 0.48 $\pm$ 0.05 & ... & 0.72 $\pm$ 0.29 & ... & D \\
2004 Oct 14.6 & 2.71 $\pm$ 0.05 & ... & ... & ... & A \\
2004 Nov 21.6 & 1.46 $\pm$ 0.06 & 1.37 $\pm$ 0.18 & < 0.21 & ... & A \\
2009 Feb 3.3 & < 0.05 & ... & ... & ... & B 
\enddata
\label{table:2004cc}
\end{deluxetable}

\begin{deluxetable}{crrrrc}
\tablecaption{Observations of SN 2004dk}
\tablehead{
\colhead{Date} & \colhead{$F_{\nu,4.9~\rm GHz}$} & \colhead{$F_{\nu,8.5~\rm GHz}$} & \colhead{$F_{\nu,15~\rm GHz}$} & \colhead{$F_{\nu,22.5~\rm GHz}$} & \colhead{VLA} \\
\colhead{(UT)} & \colhead{(mJy)} & \colhead{(mJy)} & \colhead{(mJy)} & \colhead{(mJy)} & \colhead{Config.}
}
\startdata
2004 Aug 7.1 & ... & 1.30 $\pm$ 0.05 & ... & ... & D \\
2004 Aug 11.0 & 0.87 $\pm$ 0.06 & 2.11 $\pm$ 0.06 & ... & 2.78 $\pm$ 0.33 & D \\
2004 Aug 13.0 & 1.09 $\pm$ 0.06 & 2.42 $\pm$ 0.06 & 3.57 $\pm$ 0.26 & 2.72 $\pm$ 0.37 & D \\
2004 Aug 18.1\tablenotemark{\dagger} & 0.95 $\pm$ 0.07 & 1.76 $\pm$ 0.07 & 0.83 $\pm$ 0.23 & 1.75 $\pm$ 0.46 & D \\
2004 Sep 2.0 & 1.80 $\pm$ 0.09 & 1.56 $\pm$ 0.13 & 0.79 $\pm$ 0.21 & 0.69 $\pm$ 0.18 & D \\
2004 Sep 10.0 & 1.06 $\pm$ 0.07 & 0.81 $\pm$ 0.15 & ... & < 0.43 & A \\
2004 Sep 18.0 & 0.86 $\pm$ 0.10 & 1.19 $\pm$ 0.11 & ... & < 0.47 & A \\
2004 Oct 4.1 & 0.46 $\pm$ 0.15 & 0.41 $\pm$ 0.09 & < 0.34 & < 1.41 & A \\
2004 Oct 17.8 & 0.56 $\pm$ 0.10 & 0.47 $\pm$ 0.12 & < 0.92 & ... & A \\
2005 Feb 12.4 & < 0.26 & < 0.17 & ... & ... & BnA \\
2009 Feb 12.6\tablenotemark{*} & 0.22 $\pm$ 0.06 & ... & ... & ... & B \\
2009 Feb 24.5\tablenotemark{*} & 0.21 $\pm$ 0.03 & 0.19 $\pm$ 0.03 & ... & ... & B \\
2009 Apr 2.6 & 0.34 $\pm$ 0.07 & < 0.08 & ... & ... & B \\
2009 Apr 19.0\tablenotemark{\ddagger} & 0.21 $\pm$ 0.02 & 0.15 $\pm$ 0.02 & ... & ... & C \\
2009 Sep 19.8  & 0.28 $\pm$ 0.05 & 0.17 $\pm$ 0.02 & ... & < 0.57 & B \\
2009 Oct 26.8 & ... & ... & ... & < 0.27 & D \\
\enddata
\label{table:2004dk}
\tablenotetext{\dagger}{Poor weather on 2004 Aug 18.1.} 
\tablenotetext{*}{Flux density values reported by \citet{shv+09}.} 
\tablenotetext{\ddagger}{Observations also conducted at 1.4 GHz
  revealing a flux density of 0.18 $\pm$ 0.09 mJy.}
\end{deluxetable}

\begin{deluxetable}{crrrrc}
\tablecaption{Observations of SN 2004gq}
\tablehead{
\colhead{Date} & \colhead{$F_{\nu,1.4~\rm GHz}$} & \colhead{$F_{\nu,4.9~\rm GHz}$} & \colhead{$F_{\nu,8.5~\rm GHz}$} & \colhead{$F_{\nu,15~\rm GHz}$} & \colhead{VLA} \\
\colhead{(UT)} & \colhead{(mJy)} & \colhead{(mJy)} & \colhead{(mJy)} & \colhead{(mJy)} & \colhead{Config.}
}
\startdata
2004 Dec 16.2 & ... & ... & 2.00 $\pm$ 0.05 & ... & A \\
2004 Dec 19.3 & ... & ... & 3.00 $\pm$ 0.06 & ... & A \\
2004 Dec 24.3 & ... & 1.81 $\pm$ 0.13 & 5.61 $\pm$ 0.10 & ... & A \\
2004 Dec 27.2 & ... & 2.06 $\pm$ 0.14 & 5.68 $\pm$ 0.11 & 6.27 $\pm$ 0.22 & A \\
2005 Jan 8.2 & ... & 5.75 $\pm$ 0.09 & 6.40 $\pm$ 0.06 & 3.53 $\pm$ 0.20 & A \\
2005 Jan 15.2 & < 0.92 & 4.90 $\pm$ 0.10 & 4.47 $\pm$ 0.05 & 1.96 $\pm$ 0.17 & A \\
2005 Jan 21.1 & 0.85 $\pm$ 0.24 & 4.12 $\pm$ 0.09 & 3.43 $\pm$ 0.10 & ... & BnA \\
2005 Jan 28.1 & ... & 4.45 $\pm$ 0.10 & 2.95 $\pm$ 0.08 & 1.54 $\pm$ 0.50 & BnA\\
2005 Feb 8.2 & 1.49 $\pm$ 0.27 & 3.56 $\pm$ 0.08 & 2.40 $\pm$ 0.06 & 1.64 $\pm$ 0.15 & BnA \\
2005 Feb 27.1 & 2.62 $\pm$ 0.13 & 3.44 $\pm$ 0.12 & 1.91 $\pm$ 0.07 & ... & B \\
2005 Mar 13.0 & 2.45 $\pm$ 0.55 & ... & ... & ... & B \\
2005 Mar 20.0 & 2.75 $\pm$ 0.16 & 2.68 $\pm$ 0.12 & 1.33 $\pm$ 0.09 & 0.27 $\pm$ 0.08 & B \\
2005 Apr 13.0 & ... & 1.84 $\pm$ 0.09 & 1.15 $\pm$ 0.07 & 0.63 $\pm$ 0.18 & B \\
2005 May 21.9 & 3.21 $\pm$ 0.29 & 0.97 $\pm$ 0.09 & 0.64 $\pm$ 0.06 & < 0.66 & B \\
2005 Jun 5.8 & 2.18 $\pm$ 0.17 & 0.91 $\pm$ 0.07 & 0.59 $\pm$ 0.06 & < 0.67 & B \\
2005 Aug 7.6 & 2.62 $\pm$ 0.30 & 0.57 $\pm$ 0.30 & 0.45 $\pm$ 0.07 & < 0.55 & C \\
2005 Aug 27.6 & 2.09 $\pm$ 0.31 & 1.13 $\pm$ 0.29 & 0.41 $\pm$ 0.05 & < 0.47 & C \\
2005 Oct 1.5 & 2.71 $\pm$ 0.20 & < 0.62 & 0.35 $\pm$ 0.04 & ... & C \\
2005 Oct 27.4 & ... & < 0.48 & 0.43 $\pm$ 0.14 & ... & DnC \\
2006 Jan 2.3 & ... & ... & 0.41 $\pm$ 0.05 & ... & D \\
2006 Mar 26.9 & < 0.10 & < 0.09 & < 0.14 & < 1.17 & A \\
2009 Apr 6.0 & ... & ... & < 0.08 & ... & B \\
\enddata
\label{table:2004gq}
\end{deluxetable}

\begin{deluxetable}{c|cc|cc|cc}
\tablecaption{Model Parameter Fits}
\tablehead{
\colhead{} & \colhead{2004cc} & \colhead{} & \colhead{2004dk} & \colhead{}
& \colhead{2004gq} & \colhead{} \\
\colhead{} & \colhead{Best Fit} & \colhead{Fixed $n\propto r^{-2}$} & \colhead{Best Fit} & \colhead{Fixed $n\propto r^{-2}$} & \colhead{Best Fit} & \colhead{Fixed $n\propto r^{-2}$} 
}
\startdata
$C_f$ ($\times 10^{-52}$) & 5.0 & 7.7 & 31 & 30 & 39 & 32 \\
$C_\tau$ ($\times 10^{38}$) & 0.30 & 1.9$\times 10^{-2}$ & $6.2 \times 10^{-4}$ & $6.8 \times 10^{-4}$ & $2.4 \times 10^{-3}$ & $3.4 \times 10^{-3}$ \\
$\alpha_R$ & 0.9 & 0.9 & 0.9 & 0.9 & 0.8 & 0.9 \\
$\alpha_B$ & -1.4 & -1.0 & -1.0 & -1.0 & -0.8 & -1.0 \\
$s$ & 2.8 & 2.0 & 2.0 & 2.0 & 1.5 & 2.0 \\ 
\hline \\
$R_0$ (cm) & $2.6 \times 10^{15}$ & $2.7 \times 10^{15}$ & $5.0 \times 10^{15}$ & $5.0 \times 10^{15}$ & $6.7 \times 10^{15}$ & $6.2 \times 10^{15}$ \\
$\overline{v_0}$ (c) & 0.10 & 0.10 & 0.20 & 0.19 & 0.26 & 0.24 \\
$B_0$ (G) & 8.4 & 4.2 & 1.3 & 1.3 & 1.5 & 1.6 \\ 
$E_0$ ($\times 10^{47}$ erg) & 10 & $2.7$ & $1.7$ & $1.8$ & $5.2$ & $5.1$ \\
$n_{e,0}$ ($\times 10^4~e^-$ / $\rm{cm}^3$) & $170$ & $83$ & $2.2$ & $2.3$ & $1.7$ & $2.2$ \\
$\gamma_{m,0}$ & 1 & 1 & 1.8 & 1.8 & 3.0 & 2.9 \\
$\dot{M}_0$ ($\times 10^{-5}~\rm{M}_\odot~yr^{-1}$) & $13$ & $6.5$ & $0.60$ & $0.63$ & $0.83$ & $0.92$ \\
\hline \\
Modulation Epoch  & 40,135 &  \nodata & 1708 & \nodata & 474 & \nodata \\
Abruptness Factor & 0.27,1.2 & \nodata & 7.7 & \nodata & 0.21 & \nodata \\
Modulation Factor & < 2.5,11 & 2.9 & 40 & 53 & < 0.20 & < 0.28 

\enddata
\tablecomments{The subscript ``0'' denotes the parameter values at $\Delta t\approx 10$ days since explosion.}
\label{tab:master}
\end{deluxetable}

\end{document}